\documentstyle[preprint,aps,prb,epsf,floats]{revtex}

\begin{document}

\draft


\title{Current-induced conductance switching in epitaxial 
[La$_{0.7}$Sr$_{0.3}$MnO$_3$/SrTiO$_3$] multilayers}

\author{K. D\"{o}rr, K.-H. M\"{u}ller, T. Walter, M. Sahana, D. Eckert, 
K. Nenkov, and L. Schultz}
\address{Institute of Solid State and Materials Research Dresden,
P.O.B. 270016, 01171 Dresden, Germany}
\author{K. Brand and M. Lehmann}
\address{Institute of Applied Physics, University of Technology, 
01062 Dresden, Germany}

\date{\today}

\maketitle

\begin{abstract}
We report on the non-linear in-plane electrical transport in coherently grown 
[La$_{0.7}$Sr$_{0.3}$MnO$_3$/SrTiO$_3$] multilayers with ultrathin ($<$ 3 nm) 
single layers. Current-induced switching of the conductance, with {\it low}
conductance at larger currents, is demonstrated. The conductance switching is
modified under a magnetic field, resulting in an extremely large 
magnetoresistance of negative, or in a special case even positive, sign. 
Our results suggest a percolative nature of transport where a large local 
current density gives rise to a spin-polarizing and a thermal effect of current.
\end{abstract}

\pacs{PACS numbers: 75.30.Vn, 75.70.Cn, 71.30+h, 73.50.Fq}


\narrowtext

The narrow conduction band of metallic
manganites R$_{1-x}$A$_x$MnO$_3$ (R = La or rare earth metal, A = di- or 
monovalent metal, x $\sim$ 0.3) shows a very high spin polarization at the 
Fermi energy \cite{coe,par}. Therefore, pronounced phenomena of spin-polarized 
transport can be observed, including a very large tunneling 
magnetoresistance in ferromagnet-insulator-ferromagnet junctions \cite{jo2}.
The double exchange (DE) model successfully explains the 
associated appearance of metallicity and ferromagnetism in this class of 
materials although it is not sufficient to elucidate the complexity of phases 
because any coupling of charge or spin to the lattice (e.g. the 
strong Jahn-Teller effect) is neglected \cite{coe}.  Low doping x or lattice 
distortions lead to competition of DE mediated by itinerant electrons with 
antiferromagnetic superexchange interactions of localized Mn 3d states. These 
compounds tend to be antiferromagnetic or spin-canted insulators, but some of 
them have been proven to show phase separation into metallic ferromagnetic and 
insulating clusters \cite{coe,ueh,fae,raq,doe}. Additional to charge 
localization, some insulating manganites show the characteristic feature of 
charge ordering, in particular for a commensurate fraction of charge carriers 
(e. g., x = 0.5). 

Another scenario of competing double exchange and superexchange is found at 
surfaces of ferromagnetic manganites \cite{cal}. In the first two or three 
layers of perovskite cells (of $\sim$3.9 {\AA} height), reduced lattice 
symmetry at the surface suppresses the DE. This probably causes the 
semiconducting nature of ultrathin ($\le$ 2 nm) manganite films, even if 
epitaxial strain is very low (0.1\%) \cite{su1,wa1}. Thin films can be 
positioned in an epitaxial multilayer, where their lattice structure and 
magnetization can be evaluated more precisely.

Recent work on charge-ordered Pr$_{1-x}$Ca$_x$MnO$_3$ (x = 0.3 to 0.5) 
crystals \cite{kir,fie} and charge-ordered thin films of several x = 0.5 
compounds \cite{osh,pon} has demonstrated the induction of metallic 
conductivity by application of an electric field or by creation of carriers 
via irradiation with light or x-rays \cite{kir}. A metallic path develops 
through the charge-ordered background, in some cases showing switching to 
high conductance at a threshold voltage. On the other hand, a {\it spin} 
current might well have an effect on the magnetic order of localized Mn 
spins (for theoretical description of this interaction see e.g. Ref.15). 
Indeed, there are experiments which have been interpreted in this way 
\cite{su2,wes}. They are characterized by a current of high density between 
ferromagnetic regions of manganite(grains, or particles in a tunnel junction). 

In this work we report on the non-linear {\it in-plane} electrical transport of 
coherently grown [La$_{0.7}$Sr$_{0.3}$MnO$_3$/SrTiO$_3$] multilayers  
with ultrathin manganite layers. The results indicate a phase separation into 
metallic and insulating clusters and percolative transport in the ultrathin 
La$_{0.7}$Sr$_{0.3}$MnO$_3$ films. Near the critical thickness of the 
manganite layers below which metallic conduction is lost, a current-induced 
switching of the conductance is observed. Additional application of a 
magnetic field leads to an extremely large, discontinuous, negative or even 
positive magnetoresistance in a certain range of the parameters temperature 
and current. Systematic dependence of the transport behavior on the thickness 
of manganite layers has been found. The observations are discussed within the 
framework of a spin-polarizing and a thermal effect of the current.   

The [La$_{0.7}$Sr$_{0.3}$MnO$_3$/SrTiO$_3$]$_{20}$ multilayers used in the 
experiments have been grown by off-axis pulsed laser deposition as described 
previously \cite{wa2}, using a substrate temperature of 825 $^o$C and a 
deposition rate of 1 {\AA} s$^{-1}$. The lattice mismatch of the insulating 
SrTiO$_3$ is +0.9 \%. From the numerous satellite peaks in standard x-ray 
diffractograms the superlattice period of the samples of 4.0 nm $\le$ 
$\Lambda$ $\le$ 6.35 nm has been determined. Thickness of manganite layers has
been varied between 1.6 nm (about 4 perovskite cells) and 2.9 nm (about 7 
cells). Images of the sample cross sections obtained from high resolution 
transmission electron microscopy (Fig.~\ref{Fig.1}) prove a coherent film 
growth. Some local thickness variations by one unit cell are present, as well 
as a certain curvature of the layers, that increases from substrate to surface 
of the multilayers. Resistance and magnetoresistance have been measured in 
current-in-plane (CIP) geometry, using four in-line contacts made of silver 
paint, with the magnetic field oriented along the current direction. Current-
voltage (I-V) curves were registered  at several temperatures and magnetic 
fields using a current source.  Magnetization has been determined in a SQUID 
magnetometer, including its dependence on a dc current flowing between the two 
sample ends.  

In contrast to the well-known metallic behavior of thicker epitaxial 
La$_{0.7}$Sr$_{0.3}$MnO$_3$ films, the macroscopically observed in-plane 
resistivity $\rho$$_{eff}$ of the present multilayers shows a complicated 
dependence on the measuring current I (examples shown in Fig.~\ref{Fig.2}). 
With decreasing thickness d of the manganite layers, the overall character of 
$\rho$$_{eff}$(T) changes, as expected, from metallic (Fig.~\ref{Fig.2}a) to 
semiconducting (Fig.~\ref{Fig.2}b). The non-linear behavior can be 
characterized as follows: For low currents, the resistance decreases with 
increasing current.  Thus, 
the enhancement of the conductance is a consequence of current flow. 
However, for currents above a threshold value, the resistance {\it increases}.
This is also obvious in the I-V characteristics (example in Fig.~\ref{Fig.3}),
where a hysteretic switching towards higher voltage is seen at a
switching current I$_0$, after a gradual increase of the differential 
conductance dI / dV below I$_0$.  
Striking jumps have also been observed in $\rho$$_{eff}$(T) curves (Fig.2). 
The jump temperature, T$_0$, always shifts with increasing current 
in a way that the high-resistance part of the $\rho$$_{eff}$(T) curve 
expands. (For instance, in Fig.~\ref{Fig.2}a, T$_0$ moves down with increased 
current, while in Fig.~\ref{Fig.2}b it shifts upwards.) 

The jumps are well reproducible if the sample is kept well below 
the ferromagnetic ordering temperature, T$_C$. As an example, 
Fig.~\ref{Fig.2}b shows the same $\rho$$_{eff}$(T) jump measured for the 
two directions of 
current, with +I during cooling and -I during warming. After heating above 
T$_C$ ($<$ 300 K) or
change of contact positions at room temperature, slight shifts of I$_0$
(in the range of 20 \%) and T$_0$ (by few K) have been registered. \cite{foot}

The Curie temperature of the samples reduces with decreasing 
d, from T$_C$ = 260 K for d = 2.9 nm to 165 K for 1.9 nm, and the saturation 
magnetization (44 to 38 emu g$^{-1}$) is much lower than expected for the 
Mn spins being collinear (91 emu g$^{-1}$). 
In order to check the magnetic response to current, the magnetization M has 
been measured under applied current. 
Within an accuracy of better than 0.2 \%, there was no systematic change of 
the magnetization in dependence on the applied current (not presented in a 
figure). In particular, no anomalies of M are associated with the conductance 
jumps. This result rules out a magnetic coupling phenomenon of adjacent 
manganite layers  \cite{slo}, as well as a phase transition, since both would 
be reflected in a change of magnetization. Thus, the striking variations of 
conductance do not significantly affect the magnetic order of the sample 
{\it volume}. Instead, assuming a magnetic phase separation for the ultrathin 
manganite layers, as it is also suggested for the interface region by recent 
tunneling experiments \cite{jo2}, the electrical transport seems to be 
dominated by metallic paths within the manganite layers. These paths will be 
interrupted by nm-size high-resistance "barriers" if the thickness of the 
manganite layers is slightly below the value where percolation appears. For the
current-induced {\it increase} of conductance, an aligning of Mn spins within 
the barriers, mediated by a spin-polarized current of high density seems likely. 
Also, magnetic coupling of ferromagnetic clusters via the spin current 
\cite{hei,slo} might be involved. Current-dependent switching of a manganite 
cluster surrounded by ferromagnetic manganite layers has been described in that 
way recently \cite{su2}. The microscopic origin of the observed polarizing 
effect of current needs further clarification.

In the following we suggest a mechanism explaining the effect of larger 
currents to reduce the conductance. 
Assuming a dominating role of small barriers (with reduced magnetic order
resulting in increased resistivity) within the conduction path, the thermal 
energy produced by the current at these barriers might be substantial. 
The temperature profile within the manganite layers depends on the 
distribution of both, the dissipated energy density and the thermal 
conductivity; furthermore, it also depends on the substrate temperature. 
When either heat dissipation becomes large enough or thermal conductivity 
becomes low enough that the local temperature approaches the (local) T$_C$, 
the resistivity of the considered barrier strongly increases. As an example, 
for a local resistivity of 10 $\Omega$cm within a barrier of (2.5 nm)$^3$ in 
extension and a thermal conductivity of 1 W K$^{-1}$ m$^{-1}$ for the 
manganite \cite{hej}, a 
current of 500 nA would produce a temperature difference of $\sim$40 K to 
the environment. Therefore, the thermal effect is unlikely in single films:
The thermal resistance from the barrier to the sample surface (typically 
being at He atmosphere during measurement) would be much smaller. 
Furthermore, the thermal effect of current shows an inherent tendency to 
produce switching of the conductance, being well-known as "thermal switching" 
\cite{bur}. The strong rise of the resistance of metallic manganites when T$_C$ 
is approached from below \cite{coe} might well result in a process where a 
cluster switches from the low-resistance metallic state to the high-resistance 
semiconducting state due to Joule microheating. The existence of threshold 
values for the current or the sample temperature can be expected. The magnitude 
of the resulting resistance jump will depend on how much the concerned switching 
barrier contributes to the resistance. The mechanism of thermal switching 
provides a consistent explanation of the discontinuous changes of conductance 
in our data. In particular, the $\rho$$_{eff}$ jump observed at 55 K in 
Fig.~\ref{Fig.2}b can be ascribed to the reduced thermal conductivity at low 
temperatures. It is important to note that, in contrast to the current-induced 
formation of a metallic path, the thermal effect is always characterized by 
d$\rho$$_{eff}$/dI $>$ 0.

Now we turn to the results of transport in an external magnetic field H. 
In the whole temperature range, from 5 K to 250 K, a large magnetoresistance 
(MR) in a field of 5 T is observed, as can be expected from the reduced magnetic 
order in the manganite layers. Concerning the field dependence of the 
transition temperatures, an interesting observation is made: While the 
metal-insulator transition (if any) shows the usual shift towards higher 
temperatures in a magnetic field, the temperature T$_0$ of the conductance 
jump {\it decreases}. Consequently, a large {\it positive} MR 
is found in a range below T$_0$ for metallic-like samples (Fig.~\ref{Fig.2}a). 
The MR for this peculiar case will be described later.
In the high-resistance state, the observed negative MR shows a hysteretic
switching (Fig.~\ref{Fig.4}a) in a certain range of current. 
Most measured resistance loops R(H) are reproducible for repeated field 
cycling, e. g. the loop in Fig.~\ref{Fig.4}a, where the resistance reproducibly 
switches by a factor of 8 in a field of 20 mT. 
Moreover, switching fields H$_S$ increase with current (dH$_S$ / dI $>$ 0, 
compare also I-V curves under magnetic field, Fig.~\ref{Fig.3}). This agrees 
with the above discussed general rule that larger currents stabilize the high-
resistance state. The latter property is also found for the striking positive
MR obtained just below T$_0$ for metallic-like samples (Fig.4b). The hysteretic
switching towards large resistance in high magnetic field has been observed 
for several samples at appropriate current. As expected, it is characterized 
by dH$_S$ / dI $<$ 0, in contrast to the switching with negative MR. The strong
sensitivity to current is worth noticing: H$_S$ varies by more than 1 T at a
10 \% change of I. The function of H$_S$(I) is roughly linear for both types 
of switching, with positive or negative MR.

The discussion of the striking field effect on the non-linear transport 
remains hypothetical at present. It is known that the magnetic field enhances 
the magnetic order of the Mn core spins {\it everywhere throughout the sample 
volume}. Thereby, an insulator-to-metal transition can be induced, 
without any additional effect of current \cite{det}. (In this case, 
transition fields should be insensitive to current.)  Also for the non-linear 
transport,  the changes of size and distribution of the metallic phase brought 
about by the field \cite{fae} will be essential.  Probably, the local current 
density j at the conductance path will be reduced due to the increased 
magnetic order. Thereby, a thermal effect being proportional to j$^2$ would 
be more suppressed than a polarizing one.  This might contribute to the 
stabilization of the low-resistance state in a magnetic field (case of Fig.4a).
However, the origin of the positive MR and the downshift of T$_0$ in the field 
must be different. It can be speculated that the electrical transport in a 
multilayer in a large magnetic field resembles that of another multilayer 
with somewhat thicker manganite layers in zero field, since the effect of 
the external field adds to the ferromagnetic double exchange. This very 
simple idea is supported by experimental results: Both, increase of 
d and application of a magnetic field typically lead to a drop of T$_0$ and 
to larger values of I$_0$. However, the complex interplay of current and 
field effects on the conduction process certainly needs further investigation. 

In conclusion, we have  observed non-linear in-plane electrical transport 
in coherently grown [La$_{0.7}$Sr$_{0.3}$MnO$_3$/SrTiO$_3$] multilayers 
with ultrathin single layers. Current-induced switching of the conductance, with
low conductance at larger currents, has been demonstrated. Furthermore, 
large discontinuous changes of the conductance also appear in dependence on 
temperature and magnetic field. Our results indicate a magnetic phase separation
within the ultrathin manganite layers and a percolative nature of transport. 
Non-linear transport is related to a large local current density. The latter is
suggested to give rise to a spin-polarizing and a thermal effect, which favor
a metallic and a high-resistance state, respectively.
 
We acknowledge valuable discussion with U. K. R\"{o}{\ss}ler and some 
transport measurements by M. Doerr. This work was supported by the DFG, 
SFB 422.

\newpage

\begin{figure}
\caption{HRTEM image of a section near the surface of a 
[La$_{0.7}$Sr$_{0.3}$MnO$_3$(M)/SrTiO$_3$(S)]$_{20}$ multilayer 
with single layers of 2.9 nm / 3.2 nm thickness. 
The curvature of the layers decreases towards the bottom of the multilayer.}
\label{Fig.1}
\end{figure}

\begin{figure}
\caption{
Effective in-plane resistivity $\rho$$_{eff}$ vs temperature T 
of [La$_{0.7}$Sr$_{0.3}$MnO$_3$/SrTiO$_3$]$_{20}$ multilayers 
measured during warming at different dc currents I. 
Layer thicknesses in nm are (a) 2.3/2.7 and (b) 1.9/2.3.
Lines represent straight connections of measuring points.
T$_0$ is the (current-dependent) temperature of the conductance jump.
At low currents, $\rho$$_{eff}$ shows {\it time-dependent fluctuations}.
Inset of (b): Voltage response for the two opposite directions 
of I.}
\label{Fig.2}
\end{figure}

\begin{figure}
\caption{In-plane current (I) - voltage (V) characteristics of a 
[La$_{0.7}$Sr$_{0.3}$MnO$_3$(1.6 nm)/SrTiO$_3(1.6 nm)$]$_{6}$ multilayer, 
measured at several longitudinal magnetic fields.
Such characteristics are typical for samples and temperatures where negative
magnetoresistance of switching type is found.}
\label{Fig.3}
\end{figure}

\begin{figure}
\caption{In-plane resistance vs. magnetic field for two 
[La$_{0.7}$Sr$_{0.3}$MnO$_3$/SrTiO$_3$]$_{20}$ samples with layer thicknesses
in nm of (a) 1.9/2.3 and (b) 2.9/3.2.  
After zero field cooling, the field sweep was 
$\mu_0$H = 0 $\rightarrow$ 5 T (1.) $\rightarrow$ 0 (2.) $\rightarrow$ 
-5 T (3.) $\rightarrow$ 0 (4.).
(a) Negative magnetoresistance (MR) in the high-resistance state.
(b) Positive MR observed at certain temperatures in the low-resistance state 
(see text), for different measuring currents. The 12 $\mu$A curve has 
been shifted by +50 $\Omega$ for clarity.}
\label{Fig.4}
\end{figure}

\end{document}